\documentclass[prd,preprint,preprintnumbers,nofootinbib,eqsecnum,superscriptaddress]{revtex4}

 \usepackage[dvips,final]{graphicx}
  \usepackage{amssymb}
   \usepackage{amsmath}
    \usepackage{amsfonts}
     \usepackage{epsfig}
      \usepackage{bm}% bold math

\usepackage{mathpazo}

\usepackage[section]{placeins}

\usepackage{multirow}
\usepackage{ctable}
\usepackage{booktabs}
\usepackage{array}
\usepackage{tabularx}
\usepackage{xcolor}
\usepackage{pstricks}
\definecolor{fred}{rgb}{0.90053, 0.00369, 0.00159}  % ta3skyblue

\newcommand{\GeV}{\textrm{GeV}}

\begin{document}

\author{Antoni Szczurek\footnote{also at University of Rzesz\'ow, PL-35-959 Rzesz\'ow, Poland}}
\email{antoni.szczurek@ifj.edu.pl} 
\affiliation{Institute of Nuclear
Physics, Polish Academy of Sciences, ul. Radzikowskiego 152, PL-31-342 Krak{\'o}w, Poland}

\title{
Production of axial-vector mesons at $e^+ e^-$ collisions with double-tagging
as a way to constrain the axial meson LbL contribution 
to muon g-2 and/or hyperfine splitting of muonic hydrogen}

\begin{abstract}
We calculate cross sections for production of axial-vector $f_1(1285)$
mesons for double-tagged measurements of the $e^+ e^- \to e^+ e^- f_1(1285)$
reaction. Different $\gamma^* \gamma^* \to f_1(1285)$ vertices from
the literature are used. Both integrated cross section as well as differential
distributions are calculated. Predictions for a potential measurement
at Belle II are presented. Quite different results are obtained
for the different vertices proposed in the literature.
Several observables are discussed. The distribution in photon virtuality
asymmetry is especially sensitive to the $\gamma^* \gamma^* \to f_1$ vertex.
Future measurements at $e^+ e^-$ colliders could test and/or constrain
the $\gamma^* \gamma^* \to f_1 (a_1, f_1')$ vertices and associated 
form factors, known to be important ingredients for calculating 
contributions to anomalous magnetic moment of muon and hyperfine 
splitting of levels of muonic atoms.
\end{abstract} 

\maketitle

%----------------------------
\section{Introduction}
%----------------------------

The coupling of neutral mesons to two photons is an important ingredient
of mesonic physics. In Ref.\cite{KWZ1974} tensorial coupling
was discussed for different types of mesons (pseudoscalar, scalar,
axial-vector and tensor). In general, the amplitudes can be expressed
in terms of functions of photon virtualities often called
transition form factors. They were tested in details for pseudoscalar
mesons ($\pi^0$, $\eta$, $\eta'$). Recently there was discussion how to
calculate such objects for peseudoscalar \cite{BGPSS2019} and scalar
\cite{BPSS2020} quarkonia.

The axial vector mesons and in particular their coupling
to photons \cite{PPV2012} are very important in the context of 
their contribution to anomalous magnetic moment of muon 
\cite{PV2014,OPV2017,RS2019,DMMRZ2019,LR2019,CCAGI2020}.

The anomalous magnetic moment of muon is one of the most fundamental
quantities in particle physics (see e.g. \cite{JN2009,Jegerlehner2017}).
A first calculation of QED corrections to anomalous magnetic moment
was performed long ago \cite{Schwinger1948}.
Recent state of art can be found e.g. in \cite{JN2009,Jegerlehner2017,A2020}.
The current precision of QED calculation is so high that hadronic
contributions to muon anomalous moment must be included.
The so-called light-by-light (LbL) contributions are very important 
but rather uncertain.
The coupling $\gamma^* \gamma^* \to f_1(1285)$ is one of the most
uncertain ingredients.
Different couplings have been suggested in the literature.

Recently the contribution of the $\gamma^* \gamma^* \to f_1(1285)$ coupling
was identified and included in calculating hyperfine
splitting of levels of muonic hydrogen, and turned out to be quite
sizeable \cite{DKMMR2018}.
These are rather fundamental problems and better contraints on
$\gamma^* \gamma^*$ coupling are badly needed.

In calculating $\delta a_{\mu}^{f_1}$ one often writes:
\begin{equation}
\delta a_{\mu}^{f_1} = \int d Q_1^2 d Q_2^2 
\;\; \rho_{\mu}^{f_1}(Q_1^2,Q_2^2)  \; ,
\label{density_of_anomalous_moment}
\end{equation}
where $\rho_{\mu}^{f_1}(Q_1^2,Q_2^2)$ is the density of the $f_1$
contribution to the muon anomalous magnetic moment.
The integrand of (\ref{density_of_anomalous_moment}) 
(called often density for brevity)
peaks at $Q_1^2, Q_2^2 \sim$ 0.5 GeV$^2$ and gives
almost negligible contribution for $Q_1^2, Q_2^2 >$ 1.5 GeV$^2$, see
e.g. \cite{DMMRZ2019}.

The $\gamma^* \gamma^* f_1(1285)$ coupling can be also quite important
for hyperfine splitting of levels of muonic hydrogen \cite{DKMMR2018}.
It is also very important to calculate rare decays 
such as $f_1(1285) \to e^+ e^-$ \cite{Rudenko2017,MR2019}.
There both space-like and time-like photons enter corresponding
loop integral(s) so one tests both regions simultaneously.
The corresponding branching fraction is very small ($BF \sim$ 10$^{-8}$).
%The CLAS collaboration measured such a rare decay \cite{CLAS}.
The same loop integral enters the production of $f_1$ in electron-positron
annihilation \cite{Rudenko2017,MR2019}. There is already a first
evidence of such a process from the SND collaboration at VEPP-2000
\cite{SND}.
The $f_1(1285)$ was also observed in $\gamma p \to f_1(1285) p$ reaction
by the CLAS collaboration \cite{CLAS}.
The experimental results do not agree with theoretical predictions
\cite{KBV2009,DGH2009,HXCHZ2014}.

\begin{figure}
\includegraphics[width=8cm]{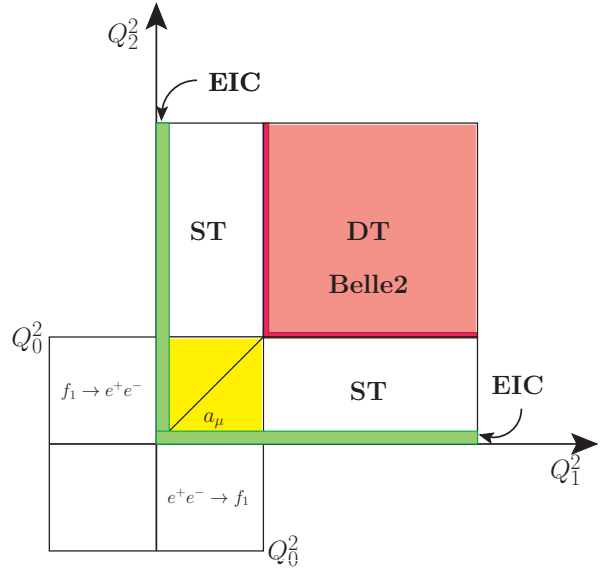}
\caption{Possible tests of the $\gamma^* \gamma^* \to AV$ vertex in
the $(Q_1^2,Q_2^2)$ space: contribution to $g-2$, hyperfine splitting
of muonic hydrogen, EIC, $f_1 \to e^+ e^-$ or $e^+ e^- \to f_1$ and 
DT in $e^+ e^-$ collisions discussed in the present paper in extent.
}
\label{fig:general_situation}
\end{figure}

Fig.\ref{fig:general_situation} illustrates how different regions of
the vertex functions are tested in different processes.
The square $(0,Q_0^2) x (0,Q_0^2)$ close to the origin shows the region
where the dominant contributions to $g-2$ comes from.
The square $(Q_0^2, \infty) x (Q_0^2, \infty)$ marked in red represents 
the region which can be tested in double-tagging experiments.
The short diagonal ($Q_1^2 = Q_2^2$) line represents region important 
for hyperfine splitting of levels of muonic hydrogen.
The narrow strips along the $x$ and $y$ axis shows a possibility
to study production of $f_1(1285)$ in $e + A$ collisions at EIC. 
Marked is also the region of photon virtualities which contributes to
$f_1 \to e^+ e^-$ or to the production of $f_1(1285)$ in $e^+ e^-$
annihilation.

%-------------------------------------------------------------
%--------------------------------------------------------------

In the present paper we suggest how to limit the behaviour of
the $\gamma^* \gamma^* \to f_1(1285)$ coupling(s)
\footnote{The same is true for other axial-vector $(a_1, f_1')$ mesons.} 
at somewhat larger photon virtualities
accessible at double-tagged $e^+ e^- \to e^+ e^- f_1(1285)$
measurements, where typically $Q_1^2, Q_2^2 > Q_0^2$ = 2 GeV$^2$.

%------------------------------------------------------------------
\section{Some details of the model calculations}
%------------------------------------------------------------------

\begin{figure}
\includegraphics[width=8cm]{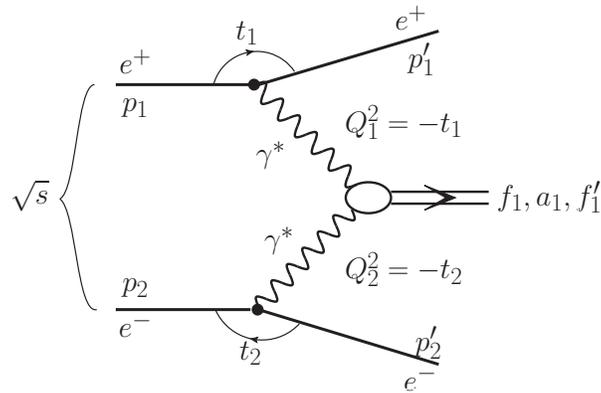}
\caption{The generic diagram for $e^+ e^- \to e^+ e^- AV$
and kinematical variables used in this paper.
}
\label{fig:diagram}
\end{figure}

Fig.\ref{fig:diagram} shows the Feynman diagram for axial-vector
meson production in $e^+ e^-$ collisions. The small circle
in the middle represent the $\gamma^* \gamma^* \to AV$ vertex
tested in double-tagging experiment.

%-------------------------------------------------------------
%--------------------------------------------------------------

%------------------------------------------------------------
\subsection{$\gamma^* \gamma^* \to f_1(1285)$ vertices}
%------------------------------------------------------------

In the formalism presented e.g. in \cite{PPV2012} the covariant 
matrix element for $\gamma^* \gamma^* \to f_1(1285)$ is written as:
where
\begin{equation}
R^{\mu,\nu} = -g^{\mu \nu} + \frac{1}{X}
\left[
(q_1 \dot q_2) \left( q_1^{\mu} q_2^{\nu} + q_2^{\mu} q_1^{\nu} \right)
- q_1^2 q_2^{\mu} q_2^{\nu} - q_2^2 q_1^{\mu} q_1^{\nu}
\right]
\end{equation}
where
\begin{equation}
X = (q_1 \dot q_2) - q_1^2 q_2^2 =
\frac{M_f^4}{4} 
\left( 1 + \frac{2(q_{1t}^2 + q_{2t}^2)}{M_f^2} + 
\frac{(q_{1t}^2 - q_{2t}^2)^2}{M_f^4} \right) \; .
\label{X}
\end{equation}

\vspace{0.5cm}

%-------------------
{\bf DKMMR2019 vertex}
%-------------------

In Ref.\cite{DKMMR2018} the vertex was written as:
\begin{eqnarray}
T^{\mu \nu}_{\alpha} = 4 \pi \alpha_{em} \epsilon_{\rho \sigma \tau
  \alpha}
&& \Big\lbrace \left.  R^{\mu \rho}(q_1,q_2)R^{\nu \sigma}(q_1,q_2)(q_1-q_2)^{\tau} \nu
F^{0}(q_1^2,q_2^2) \right. \;   \nonumber \\
&+& \left. R^{\nu \rho}(q_1, q_2) \left(q_1^{\mu} - \frac{q_1^2}{\nu} q_2^{\mu} \right) 
q_1^{\sigma} q_2^{\tau} F^{(1)}(q_1^2,q_2^2) \right.  \nonumber \\
&+& \left. R^{\mu \rho}(q_1, q_2) \left(q_2^{\nu} - \frac{q_2^2}{\nu} q_1^{\nu} \right) 
q_2^{\sigma} q_1^{\tau} F^{(1)}(q_2^2,q_1^2)  \Big\rbrace \right.   \; ,
\label{vertex_DKMMR}
\end{eqnarray}
where
\begin{equation}
\nu = (q_1 q_2) = \frac{1}{2} \left( (q_1 + q_2)^2 - q_1^2 - q_2^2 \right) \; .
\end{equation}
In the nonrelativistic model 
\begin{equation}
F^{(0)}(0,0) = -F^{(1)}(0,0)   \; .
\end{equation}
We use the normalization of form factors
\begin{equation}
F^{(0)}(0,0) = 0.266 \; GeV^{-2}  \; .
\end{equation}
In \cite{DKMMR2018} the vertex was supplemented by the following
factorized dipole form factor
\begin{equation}
F_{DKMMR}(Q_1^2,Q_2^2) = \frac{\Lambda_D^4}{(\Lambda_D^2 + Q_1^2)^2}
                         \frac{\Lambda_D^4}{(\Lambda_D^2 + Q_2^2)^2} \; .
\label{factorized_dipole}
\end{equation}
The $\Lambda_D \approx$ 1 GeV was suggested as being consistent with
the L3 collaboration data \cite{L3_f1}.

We will ascribe also the name NQM (nonrelativistic quark model)
to this vertex.

\vspace{0.5cm}

%-----------------
{\bf OPV2018 vertex}
%-----------------

In Ref.\cite{OPV2017} the vertex function for $\gamma^* \gamma^* \to f_1$
was constructed based on an analysis of the 
$f_1(1285) \to \rho^0 \gamma$ decay and vector meson dominance picture.
The corresponding vertex for two-photon coupling there reads
\begin{eqnarray}
T^{\mu \nu \alpha} = i C_{OPV} 
&&\left. \Big\lbrace
\epsilon^{\mu \nu \sigma \alpha}
(q_{1,\sigma}((q_1 q_2)+ 2 q_1^2) - q_{2,\sigma}((q_1 q_2) + 2 q_2^2))
 \right.
\nonumber \\
&+& \left. \epsilon^{\rho \sigma \nu \alpha} q_{2,\rho} q_{1,\sigma}(q_2 + 2 q_1)^{\mu}
\right.
\nonumber \\
&+& \left. \epsilon^{\rho \sigma \mu \alpha} q_{1,\rho} q_{2,\sigma}(q_1 + 2 q_2)^{\nu}
\Big\rbrace \right.
\; .
\label{OPV_vertex}
\end{eqnarray}
Above
\begin{equation}
C_{OPV} = \frac{5 \alpha_{em} g_{\rho}}{36 \pi M_{f_1}^2}   \; .
\end{equation}
The value of $g_{\rho}$ is explicitly given in \cite{OPV2017}.
We supplemented this vertex with one common for all terms form factor
of the VDM type:
\begin{equation}
F(Q_1^2,Q_2^2) = \frac{M_V^2}{M_V^2 + Q_1^2} \frac{M_V^2}{M_V^2 + Q_2^2}
\; .
\label{VDM_formfactor}
\end{equation}
consistent with the philosophy there.

\vspace{0.5cm}

%----------------
{\bf LR2019 vertex}
%----------------

\begin{figure}
\includegraphics[width=6cm]{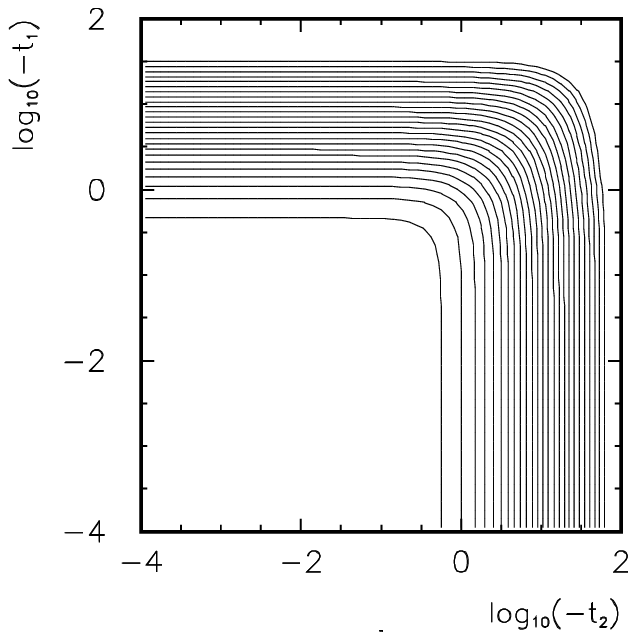}
\includegraphics[width=6cm]{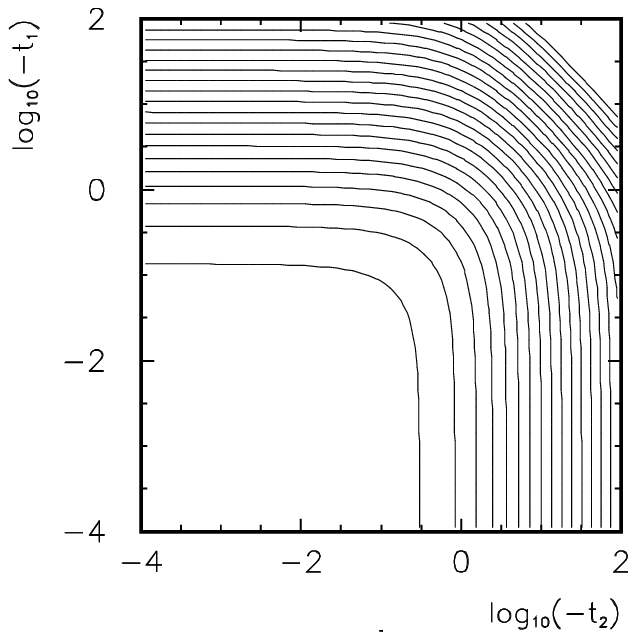}
\caption{Maps of the original (left panel) and parametrized (right panel) 
HW2 form factor $A(Q_1^2,Q_2^2)/A(0,0)$
as a function of ($\log_{10}(Q_1^2),\log_{10}(Q_2^2)$). 
In the latter case $\Lambda_S$ = 0.8 GeV and $\Lambda_L$ = 1.2 GeV.
}
\label{fig:HW2_FF}
\end{figure}

Finally we consider also the vertex used very recently in \cite{LR2019}.
In this approach the vertex is
\begin{eqnarray}
T^{\mu \nu \rho} \propto
\epsilon^{\alpha \beta \rho \sigma} 
&&\left. \Big\lbrace
  (q_{1}^2 \delta_{\alpha}^{\mu} - q_{1,\alpha} q_1^{\mu})
q_2^{\sigma} \delta_{\beta}^{\nu} A(Q_1^2, Q_2^2) \right. \nonumber \\
&& \left. - (q_{2}^2 \delta_{\beta}^{\nu} - q_{2,\beta} q_2^{\nu}) 
q_1^{\sigma} \delta_{\alpha}^{\nu} A(Q_2^2, Q_1^2)
\Big \rbrace \right. \; .
\label{amplitude_LR}
\end{eqnarray}
The normalization was also given there.
It was pointed out that the $A(Q_1^2,Q_2^2)$ function does not need to 
be symmetric under exchange of $Q_1^2$ and $Q_2^2$. Actually asymmetric 
form factors calculated from the hard wall and Sakai-Sugimoto models
were used there.
In our evalution here we will use Hard Wall (HW2) form factors as well
as factorized dipole symmetric/asymmetric form factors as specified below
to illustrate the effect of the holographic approach.
The HW2 form factor can be sufficiently well represented as:
\begin{eqnarray}
A(Q_1^2,Q_2^2) &\approx& A(0,0) F_S(Q_1^2) F_L(Q_2^2) \; , \nonumber \\ 
A(Q_2^2,Q_1^2) &\approx& A(0,0) F_L(Q_1^2) F_S(Q_2^2) \; ,
\label{HW2_parametrization}
\end{eqnarray}
where
\begin{eqnarray}
F_S(Q^2) = \left( \frac{\Lambda_S^2}{\Lambda_S^2 + Q^2 } \right)^2
\; , \nonumber \\
F_L(Q^2) = \left( \frac{\Lambda_L^2}{\Lambda_L^2 + Q^2 } \right)^2
\; ,
\end{eqnarray}
where $\Lambda_L > \Lambda_S$.
We show the HW2 form factor and its factorized dipole approximate 
representation as a function of ($\log_{10}(Q_1^2),\log_{10}(Q_2^2)$) 
in Fig.\ref{fig:HW2_FF}.

%------------------------------------------------------------------------
%------------------------------------------------------------------------

\vspace{0.5cm}

%----------------
{\bf RS2019 vertex}
%----------------

In Ref.\cite{RS2019} a vertex based on R$\chi$T approach
was considered. In this approach one gets:
\begin{eqnarray}
T^{\mu \nu \alpha} = e^2 F_{RS}(q_1,q_2)
&&\left. \Big \lbrace
i \epsilon^{\mu \tau \alpha \rho} q_{1,\rho}
(q_2^{\nu} q_{2,\tau} - g_{\tau}^{\nu} q_2^2)
-i \epsilon^{\nu \tau \alpha \rho} q_{2.\rho}
(q_1^{\mu} q_{1,\tau} - g_{\tau}^{\mu} q_1^2) \right.   \\ \nonumber
&&\left. + i \epsilon^{\mu \nu \rho \sigma} q_{1,\rho} q_{2,\sigma} 
( q_1^{\alpha} - q_2^{\alpha} ) 
\rbrace\right.   \; .
\label{RS_vertex}
\end{eqnarray}
Above we have denoted:
\begin{equation}
F_{RS}(q_1,q_2) = \frac{2 c_A}{M_A} 
\frac{(q_1^2 - q_2^2)}{(q_1^2 - M_V^2)(q_2^2 - M_V^2)} \; .
\label{RS_formfactor}
\end{equation}
The $c_A$ is defined in \cite{RS2019}.
$M_V \approx m_{\rho} \approx m_{\omega}$ = 0.8 GeV.
The reader is asked to note vanishing of $F_{RS}$
at $Q_1^2 = Q_2^2$. This, as will be discussed below,
has important consequences for the double tagged measurements.

The form factor used in RS2019 are antisymmetric.
Additional symmetric form factors arising at higher order were
discussed in a revised version of \cite{RS2019} (see Appendix C there).
In the following we will use the lower order result to illustrate 
the situation.

%In actual calculation of the cross sections for $e^+ e^- \to e^+ e^- f_1$ 
%we shall omit the last term in (\ref{RS_vertex}) which leads to 
%a violation of Bose symmetry.
It was ascertained recently in \cite{MRS2020} that the R$\chi$T approach
provides only purely transverse axial-vector meson contributions.

\vspace{0.5cm}

%----------------
{\bf MR2019 vertex}
%----------------

In Ref.\cite{MR2019} the following vertex was used (we change a bit 
notation to be consistent with our previous formulae)
\begin{eqnarray}
T^{\mu \nu \alpha} = \frac{i}{m_{f_1}^2}
\epsilon^{\mu \nu \rho \sigma} && \left. \Big\lbrace
F(q_1^2, q_2^2) q_{2 \rho} q_{1,\sigma} (q_1-q_2)^{\alpha} \right. \;
\nonumber \\
&&\left. - q_2^2 G(q_1^2, q_2^2) \delta_{\rho}^{\alpha} q_{1,\sigma}
          + q_1^2 G(q_1^2,q_2^2) \delta_{\rho}^{\alpha} q_{2,\sigma}
\Big\rbrace \right.
\label{MR2019_vertex}
\end{eqnarray}
to the production of $f_1(1285)$ in the $e^+ e^-$ annihilation.
Since in this case both space-like and time-like virtualities enter the
calculation of the relevant matrix element the form factors had to be
generalized.
In \cite{MR2019} the form factors were parametrized in the spirit of 
vector meson dominance approach as:
\begin{eqnarray}
G(q_1^2,q_2^2) &=& \frac{g_2 M_f^5}
{q (q_1^2 - m_{\rho}^2 + i m_{\rho} \Gamma_{\rho})
   (q_2^2 - m_{\rho}^2 + i m_{\rho} \Gamma_{\rho})} \; , \\
F(q_1^2,q_2^2) &=& \frac{g_1 M_f^3 (q_2^2 - q_1^2)}
{q (q_1^2 - m_{\rho}^2 + i m_{\rho} \Gamma_{\rho})
   (q_2^2 - m_{\rho}^2 + i m_{\rho} \Gamma_{\rho})} \; .
\label{MR_formfactors}
\end{eqnarray}
One can see the characteristic $\rho$ meson propagators.
The $F(q_1^2,q_2^2)$ form factor is asymmetric with respect to
$q_1^2$ and $q_2^2$ exchange to assure Bose symmetry of the amplitude.
An extra $q$ in the denominator was attached to the VDM-like vertex
to assure ``correct'' behaviour of the form factors at large photon 
virtualities \cite{KWZ1974}. Of course, it is not obvious that 
such a correction should enter in the multiplicative manner.
The coupling constant
\begin{equation}
g_2 = (2.9 \pm 0.4) \cdot 10^{-4}
\end{equation}
was found in \cite{MR2019}.
It was allowed in \cite{MR2019} for $g_2$ to be complex.
It was argue that $|g_1| \sim g_2$ to describe the first 
$e^+ e^- \to f_1(1285)$ data from VEPP-2000 \cite{SND}.
We shall show in this paper how important is the interference of both 
terms in the DT case.

\vspace{0.5cm}

%-------------------------------------------
\subsection{General requirements}
%-------------------------------------------

Any correct formulation of the $\gamma^* \gamma^* \to f_1(1285)$ vertex
must fulfill at least three general requirements:

\begin{itemize}

\item Gauge invariance requires:
\begin{equation}
q_{1 \mu} T^{\mu \nu \alpha} = q_{2 \nu} T^{\mu \nu \alpha} = 0 \; ,
\label{gauge_invariance}
\end{equation}

\item Landau-Yang theorem \cite{LY} requires:
\begin{equation}
T^{\mu \nu \alpha} \to 0 \;\;  {\rm when} \;\; q_1^2 \to 0 \; \; 
{\rm and} \; \; q_2^2 \to 0 \; .
\label{Landau_Yang}
\end{equation}

\item Bose symmetry implies
\begin{equation}
T^{\mu \nu \alpha}(q_1,q_2) = T^{\nu \mu \alpha}(q_2,q_1)
\label{Bose_symmetry1}
\end{equation}
which for our reaction means e.g.:
\begin{equation}
\frac{d \sigma (t_1,t_2;y,\phi)}{dt_1 dt_2 dy d \phi} =
\frac{d \sigma (t_2,t_1;y,\phi)}{dt_1 dt_2 dy d \phi} 
\label{Bose_symmetry}
\end{equation}
for each $y,\phi$.

\end{itemize} 

Some vertices fulfil also
\begin{equation}
T^{\mu \nu \alpha} p_{\alpha} \; ,
\end{equation}
where $p$ is four-momentum of the axial-vector meson.
This automatically guarantees that only spin-1 particle $f_1$
is involved and unphysical states are ignored.
A related discussion can be found e.g. in \cite{EMN2014}.

%-------------------------------
\subsection{Form factors}
%-------------------------------

Some of the $F(Q_1^2,Q_2^2)$ form factors can be constraint from 
the so-called decay width into transverse and longitudinal photon, some
are poorly know as they can not be obtained as they
do not enter the formula for the radiative decay width.
The radiative decay width is known \cite{PDG} and is
\begin{equation}
{\tilde \Gamma}_{\gamma \gamma} = 3.5 \ keV \; .
\end{equation}
Then some of the form factors are parametrized as:
\begin{eqnarray}
F(Q_1^2,Q_2^2) &=& \left( \frac{\Lambda_M^2}{\Lambda_M^2 + Q_1^2} \right)
                   \left( \frac{\Lambda_M^2}{\Lambda_M^2 + Q_2^2}
                   \right) \; , \\
F(Q_1^2,Q_2^2) &=& \left( \frac{\Lambda_D^2}{\Lambda_D^2 + Q_1^2} \right)^2
                   \left( \frac{\Lambda_D^2}{\Lambda_D^2 + Q_2^2}
                   \right)^2 \; , \\ 
F(Q_1^2,Q_2^2) &=& 
\left( \frac{\Lambda_M^2}{Q_{1}^2+Q_{2}^2+\Lambda_M^2} \right)  \; , \\
F(Q_1^2,Q_2^2) &=& 
\left( \frac{\Lambda_D^2}{Q_{1}^2+Q_{2}^2+\Lambda_D^2} \right)^2  \; . 
%F(q_{1t}, q_{2t}) &=& \frac{m_{\rho}^2}{m_V^2 + q_{1t}^2}
%                      \frac{m_{\rho}^2}{m_V^2 + q_{2t}^2}  \; , \\
%F(q_{1t}, q_{2t}) &=& \left( \frac{m_{\rho}^2}{m_{\rho}^2+q_{1t}^2}
%                \frac{m_{\rho'}^2}{m_{\rho'}^2+q_{1t}^2} \right)  
%                    \left( \frac{m_{\rho}^2}{m_{\rho}^2+q_{2t}^2}
%                \frac{m_{\rho'}^2}{m_{\rho'}^2+q_{2t}^2} \right) \; .      
\label{formfactors}                
\end{eqnarray}
Both monopole and dipole parametrizations of form factors will be used 
in the following. We will call the first two as factorized Ansatze and 
the next two as pQCD inspired power-like parametrizations.

In general, the form factors in Eqs.(\ref{amplitude_LR})
do not need to be symmetric with respect to $Q_1^2$ and $Q_2^2$
exchange \cite{LR2019}.
For example in Ref.\cite{LR2019} asymmetric form factor $A(Q_1^2, Q_2^2)$
obtained in Hard Wall and Sakai-Sugimoto models were used
to calculate contribution to anomalous magnetic moment of muon.
Here we shall take a more phenomenological approach and try to
parametrize the form factors in terms of simple functional forms
motivated by physical arguments such as vector dominance model
or asymptotic pQCD behaviour of transition form factors (see e.g.
\cite{DKV2001}).

The behaviour of transition form factors at asymptotia may be another
important issue \cite{HS2020}. Where the pQCD sets in is interesting but
still an open issue. It was discussed in \cite{BGPSS2019} that for 
$\gamma^* \gamma^* \eta_c$ coupling this happens at very high virtualities. 
We leave this issue for the $\gamma^* \gamma^* f_1$ coupling 
for a future study.

%----------------------------------------------------
\subsection{$e^+ e^- \to e^+ e^- f_1$ reaction}
%----------------------------------------------------

The amplitude for the $e^+ e^- \to e^+ e^- f_1$ reaction 
(see Fig.\ref{fig:diagram}) in high-energy
approximation can be written as:
\begin{equation}
{\cal M}^{\alpha} = e \left(p_1 + p_1' \right)^{\mu_1}
                  \left( \frac{i g_{\mu_1 \nu_1}}{t_1} \right)
                  T_{\gamma^* \gamma^* \to f_1}^{\nu_1 \nu_2 \alpha}
                    e \left(p_2 + p_2' \right)^{\mu_2}
                  \left( \frac{i g_{\mu_2 \nu_2}}{t_2} \right)  \; .
\label{amplitude}
\end{equation}
Above $e^2 = 4 \pi \alpha_{em}$. The four-momenta are defined in
Fig.\ref{fig:diagram}. The $T^{\nu_1 \nu_2 \alpha}$ vertex function
responsible for the $\gamma^* \gamma^* \to f_1$ coupling
was discussed in detail in the previous subsection.

The square of the matrix element, summed over polarizations of $f_1$, 
can be obtained as:
\begin{equation}
\overline{ | {\cal M} |^2 } = 
\sum_{\alpha_1,\alpha_2} {\cal M}^{\alpha_1}
                         {\cal M}^{\alpha_2}
                         P_{\alpha_1 \alpha_2}(p_{f_1}) \; ,
\label{amplitude_squared}
\end{equation}
where $P$ is spin-projection operator for spin-1 massive particle:
\begin{equation}
P_{\alpha_1 \alpha_2} = -g_{\alpha_1, \alpha_2} +
\frac{p_{\alpha_1} p_{\alpha_2}}{M_{f_1}^2}   \; .
\end{equation}

The cross section for the 3-body reaction $e^+ e^- \to e^+ e^- f_1(1285)$
can be written as
\begin{eqnarray}
d \sigma = \frac{1}{2 s} \overline { | {\cal M} |^2 } \cdot
d^{\,3} PS\, . 
\label{cross sect}
\end{eqnarray}
The three-body phase space volume element reads
\begin{eqnarray}
d^3 PS = \frac{d^3 p_1'}{2 E_1' (2 \pi)^3} \frac{d^3 p_2'}{2 E_2'
(2 \pi)^3} \frac{d^3 P_M}{2 E_M (2 \pi)^3} \cdot (2 \pi)^4
\delta^4 (p_1 + p_2 - p_1' - p_2' - P_M) \; . \label{dPS_element}
\end{eqnarray}

The phase-space for the $p p \to p p f_1$ reaction has four 
independent kinematical variables.
In our calculation we integrate over $\xi_1 = log_{10}(p_{1t})$, 
$\xi_2 = log_{10}(p_{1t})$, azimuthal angle between positron and
electron and rapidity of the produced axial-vector meson
(four-dimensional integration).
Here $p_{1t}$ and $p_{2t}$ are transverse momenta of outgoing
positron and electron, respectively.

In the case of holographic approach first the $A(Q_1^2,Q_2^2)$ form
factor entering the central vertex function (see Eq.(\ref{amplitude})) 
is calculated on a two-dimensional grid and then the grid is 
used for interpolation for each phase space point (see (\ref{cross sect})).

%-----------------------------------------
\section{Numerical predictions}
%-----------------------------------------

%-------------------------------------------------
\subsection{Low $Q_1^2$, $Q_2^2$ region}
%-------------------------------------------------

\begin{figure}
\includegraphics[width=6cm]{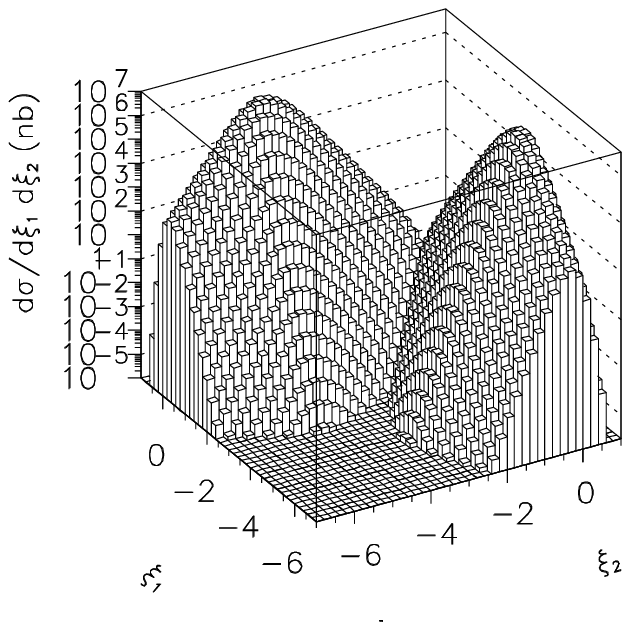}
\includegraphics[width=6cm]{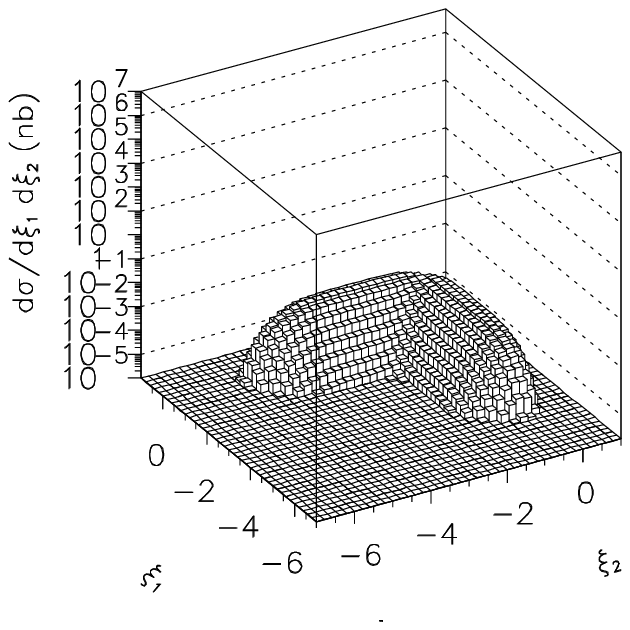} \\
\includegraphics[width=6cm]{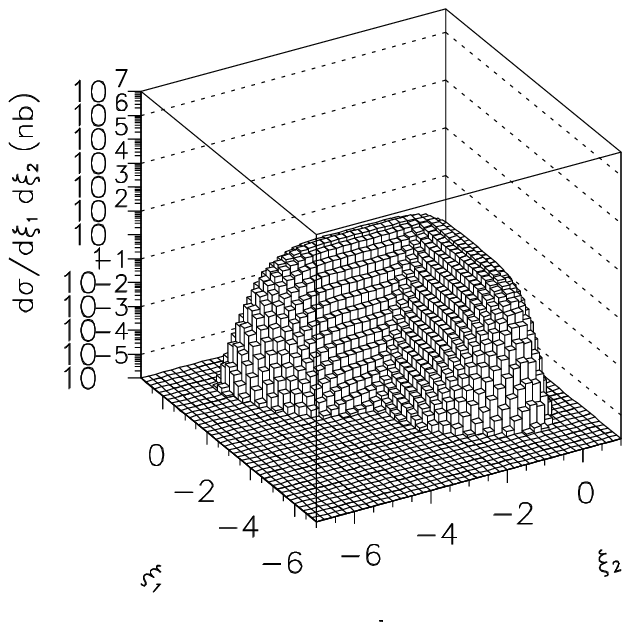}
\includegraphics[width=6cm]{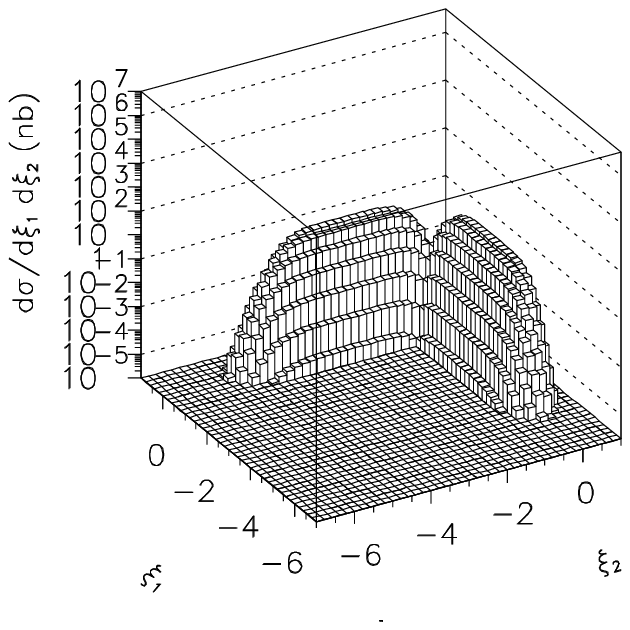}
\caption{Distributions in $\xi_1$ and $\xi_2$ for $\sqrt{s}$ = 10.5 GeV.
Here the OPV, NQM, LR and RS vertices were used.
}
\label{fig:dsig_dxi1dxi2}
\end{figure}

In Fig.\ref{fig:dsig_dxi1dxi2} we show a two-dimensional
distribution ($\xi_1, \xi_2$) of the full phase space cross section. 
Quite large cross sections are obtained for small $\xi_1$ and/or $\xi_2$.
In addition, the different models of the $\gamma^* \gamma^* f_1$
couplings lead to very different results for the total cross section.
The measurement of the total cross section is, however, rather difficult. 
%We do not get integrable cross sections for some of the couplings.

%-------------------------------------------------------------
%--------------------------------------------------------------

\begin{figure}
\includegraphics[width=6cm]{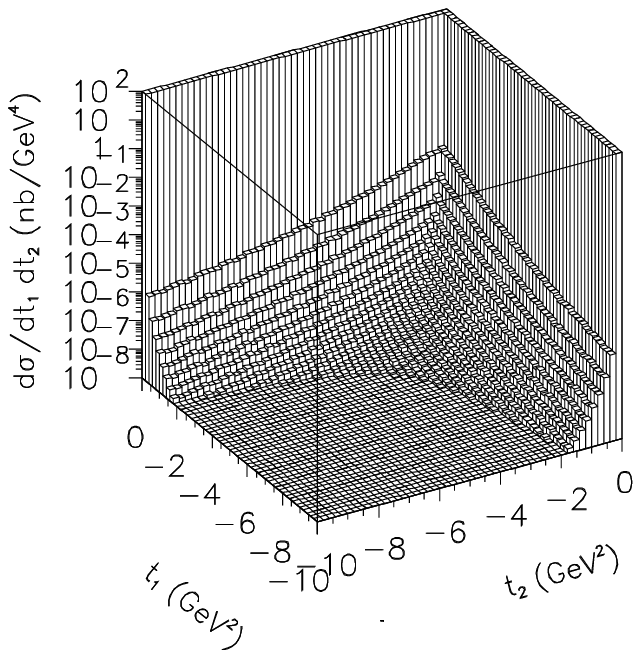}
\includegraphics[width=6cm]{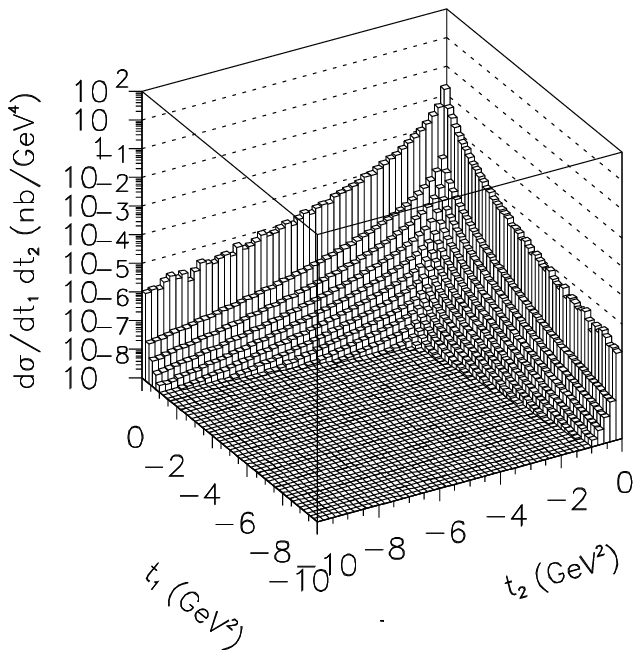}
\includegraphics[width=6cm]{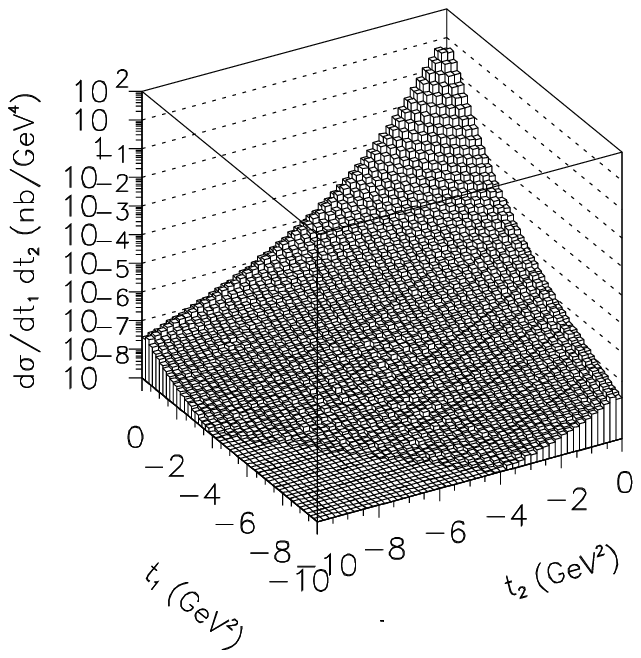}
\includegraphics[width=6cm]{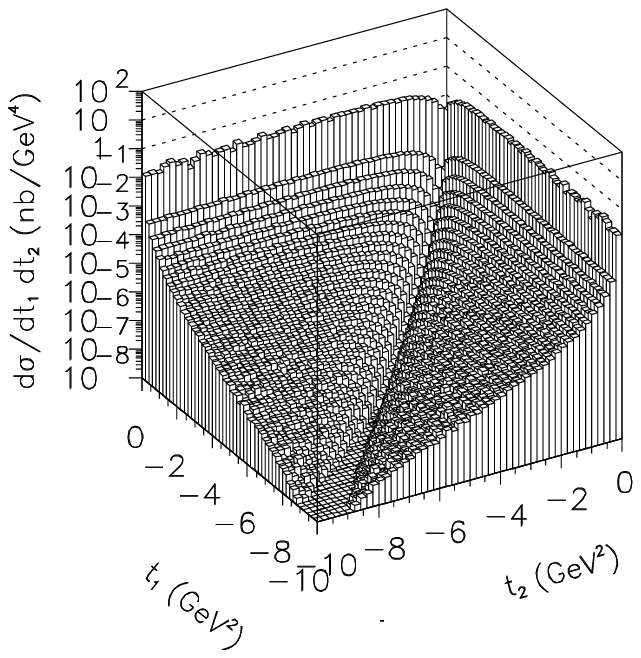}
\caption{Distributions in $t_1$ and $t_2$ for $\sqrt{s}$ = 10.5 GeV.
Here the OPV, NQM, LR and RS vertices were used.
}
\label{fig:dsig_dt1dt2}
\end{figure}

In Fig.\ref{fig:dsig_dt1dt2} we show distributions in $(t_1, t_2)$
(four-momenta squared of the virtual photons as shown in 
Fig.\ref{fig:diagram}).
Clearly some couplings generate strongly enhanced cross section at 
small $t_1, t_2$.

%-------------------------------------------------------------
%--------------------------------------------------------------

Clearly those different vertices lead to different cross sections even
for very small photon virtualities where the cross section is relatively
large. Could one measure inclusive cross section for production
of axial-vector meson without tagging ?
Is then $\gamma^* \gamma^* \to f_1(1285)$ the dominant mechanism ?
If yes, such measurements would verify the different vertices used in 
calculating $\delta a_{\mu}$ (axial-vector meson contribution to
$a_{\mu}$).
Small $Q_1^2$ and $Q_2^2$ means small transverse momenta of $f_1(1285)$.
Can one then identify $f_1(1285)$. Which channel is the best ?
This requires further Monte Carlo studies.
The resonant $e^+ e^- \to f_1(1285)$ production is very small
\cite{MR2019} and important only at resonance energies 
($\sqrt{s} \sim m_{f_1}$).
We are not aware about other competitive reaction mechanisms
in $e^+ e^-$ collisions.

\begin{figure}
\includegraphics[width=6cm]{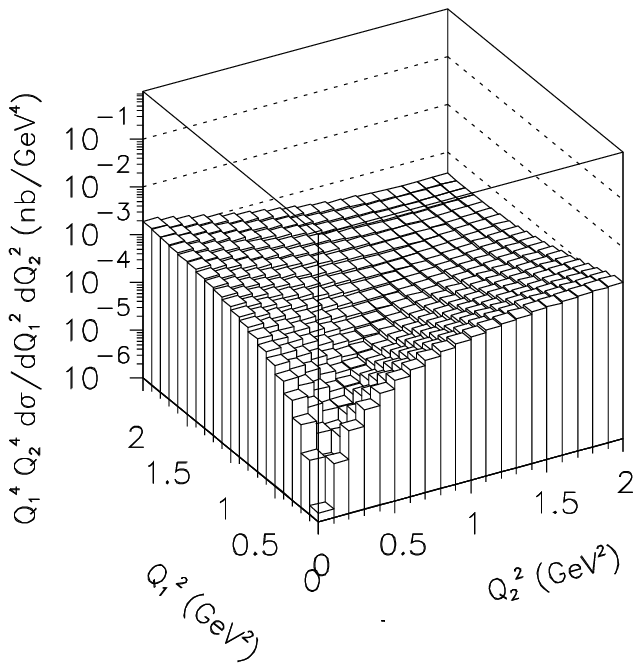}
\includegraphics[width=6cm]{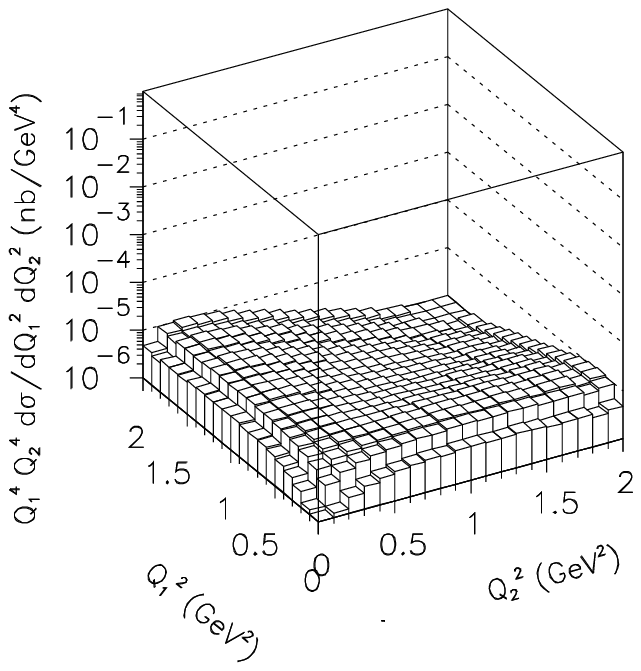}
\includegraphics[width=6cm]{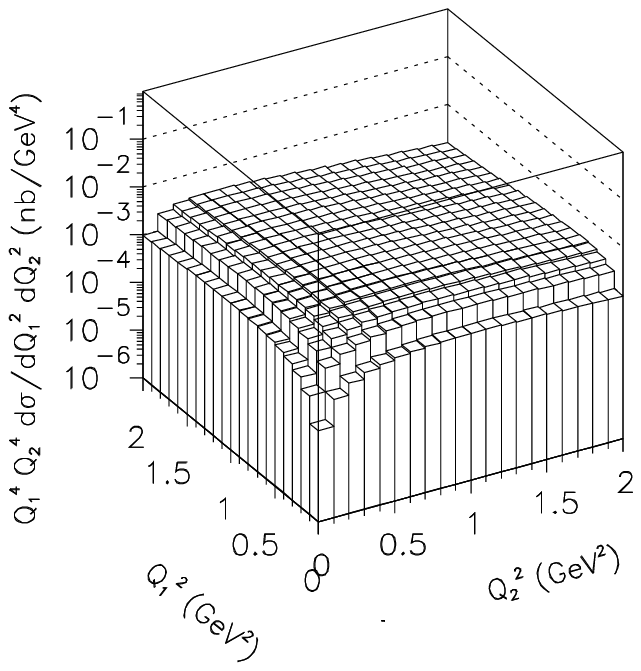}
\includegraphics[width=6cm]{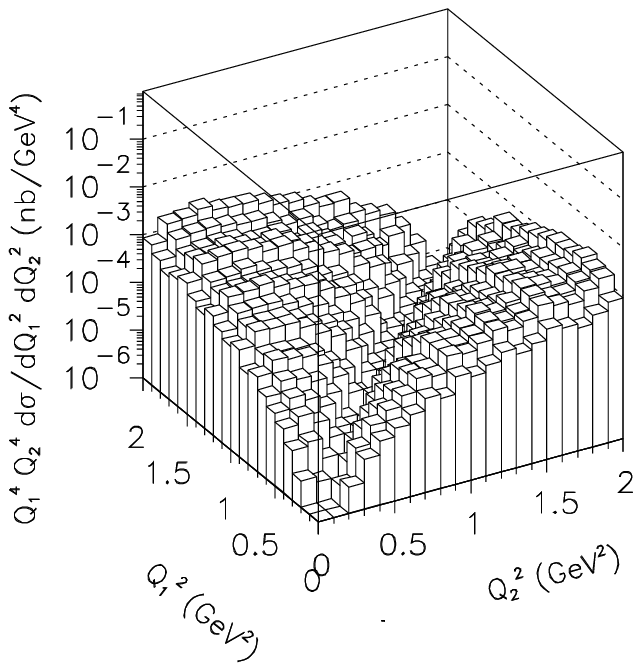}
\caption{A two-dimensional maps of the special quantity $\Omega_{LY}$ 
for $\sqrt{s}$ = 10.5 GeV.
Here the OPV (left upper), NQM (right upper), LR (left lower) and RS
(right lower) vertices were used.
}
\label{fig:Q14Q24dsig_dQ12dQ22}
\end{figure}

In general, one observes a strong enhancement of the 
$e^+ e^- \to e^+ e^- f_1(1285)$
cross section at $Q_1^2, Q_2^2 \to$ 0 which is dictated by the singular
behaviour of photon propagators in (\ref{amplitude}).
To illustrate and explore the effect of Landau-Yang vanishing of 
$T^{\mu \nu \alpha}$ vertex function for $\gamma^* \gamma^* \to f_1$ 
in Fig.\ref{fig:Q14Q24dsig_dQ12dQ22} we plot the following quantity:
\begin{equation}
\Omega_{LY}(Q_1^2,Q_2^2) = \frac{Q_1^4 Q_2^4}{M_0^4 M_0^4} 
                           \frac{d \sigma(Q_1^2,Q_2^2)}{d Q_1^2 d Q_2^2}
                           \; .
\label{special_quantity}
\end{equation}
The arbitrary scale $M_0$ is chosen to be $M_0$ = 1 GeV in the following.
%
%-------------------------------------------------------------
%--------------------------------------------------------------

One can clearly see vanishing of the special quantity 
(\ref{special_quantity}) at $Q_1^2 \to$ 0 and $Q_2^2 \to$ 0
which reflects Landau-Yang theorem. Slightly different approach
patterns to zero can be observed for the different couplings.
For the RS coupling we observe deep valley arround $Q_1^2 = Q_2^2$
which is  a direct consequence of the specific form factor used there.
In this case $\Omega_{LY}$ is much smaller than for other vertices 
in the limited range of $Q_1^2$ and $Q_2^2$ shown in the figure.

%-----------------------------------------
\subsection{Double-tagging case}
%-----------------------------------------

In Table 1 we show integrated cross sections in nb for different couplings
discussed in the previous section.
Here we imposed only Lorentz invariant cuts $Q_1^2, Q_2^2 >$ 2 GeV$^2$.
Quite different values are obtained with different couplings 
which show huge uncertainties of our predictions.
Surprisingly small cross sections are obtained with the MR2019
couplings, where we show results with different sign of the second term.
Therefore we show also contributions of individual terms for some
couplings from the literature.
They give contributions of similar order of magnitude.

%-----------------------------------------------------------------------
\begin{table}
\caption{Integrated cross section in nb for the double-tagging case with
$Q_1^2, Q_2^2 > 2 \GeV^{2}$. The MR+, MR- below show the effect of
interference due to sign changing of a ``subleading'' contribution.
%In the case of RS2019 vertex we show results for two and three terms in (\ref{RS_vertex}).
}
\begin{center}
\begin{tabular}{|c|c|c|}
\hline
vertex      &   cross section  &  comment   \\
\hline
LR          &    0.6892(-04)  &  fact. dipole, $\Lambda = $ 1 GeV   \\
            &    0.3715(-04)  &  HW2 form factor \\
OPV         &    0.9212(-04)  &  pQCD dipole,  $\Lambda = M_{f_1}$   \\
%NQM         &    0.7823(-04)  &  factorized dipole $\Lambda = $ 1 GeV \\
NQM         &    0.4905(-07)  &  factorized dipole $\Lambda = $ 1 GeV \\
%RS          &    0.2903(-02)  &  antisymmetric form factor, $\Lambda =$ 0.8 GeV, two terms\\
%            &    0.1376(-02)  &  antisymmetric form factor, $\Lambda =$ 0.8 GeV, three terms \\
RS          &    0.2138(-02)   &  antisymmetric form factor, $\Lambda =$ 0.8 GeV \\
MR +        &    0.4327(-07)   &  symmetric and antisymmetric form factors \\
MR -        &    0.7410E(-07)   &  symmetric and antisymmetric form factors \\
MR first    &    0.3432E(-07)   &  antisymmetric form factors\\      
MR second   &    0.2435E(-07)   &  symmetric form factor \\
\hline
\end{tabular}
\end{center}

\end{table}
%------------------------------------------------------------------------

The results are also strongly dependent on the form factor used in 
the calculation which is discussed below.
In Table 2 we show integrated cross section for a simple LR2019 coupling
\cite{LR2019} supplemented by the pQCD or factorized dipole form factor 
with different values of the form factor parameter $\Lambda$.
The results dramatically depend on the value of $\Lambda$.
In addition for the same $\Lambda$ the pQCD and factorized dipole Ans\"atze
give cross section for double tagged case differing by an order of magnitude.
In contrast for single tagged case they give almost the same result.

%---------------------------------------------------------------------------------------------------
\begin{table}
\caption{Integrated cross section in nb
for $e^+ e^- \to e^+ e^- f_1(1285)$ at $\sqrt{s}$ = 10.5 GeV for 
the vertex used in \cite{LR2019} for arbitrarily changed form factors. 
We present results for different values of form factor parameter.}
\begin{center}
\begin{tabular}{|c|c|c|c|}
\hline
 pQCD dipole $\Lambda$ (GeV) & $\sigma$ (nb) & factorized dipole $\Lambda$ (GeV) & $\sigma$ (nb) \\
\hline
   0.8       &  0.4477(-3) &     0.8      &  0.4292(-5)  \\ 
   1.0       &  0.2236(-2) &     1.0      &  0.6892(-4)  \\
   1.2       &  0.7867(-2) &     1.2      &  0.5432(-3)  \\
\hline
\end{tabular}
\end{center}

\end{table}
%----------------------------------------------------------------------------------------------------

\begin{figure}
\includegraphics[width=6cm]{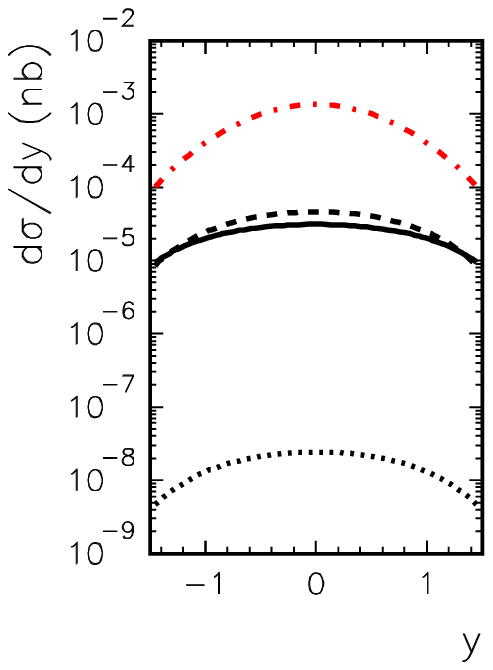}
\includegraphics[width=6cm]{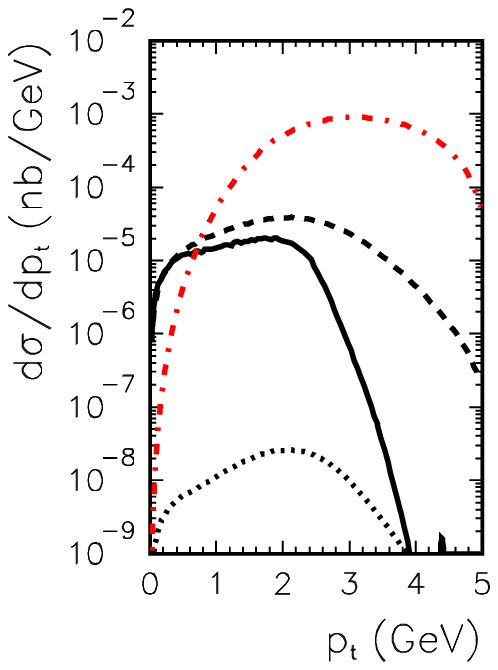} \\
\includegraphics[width=6cm]{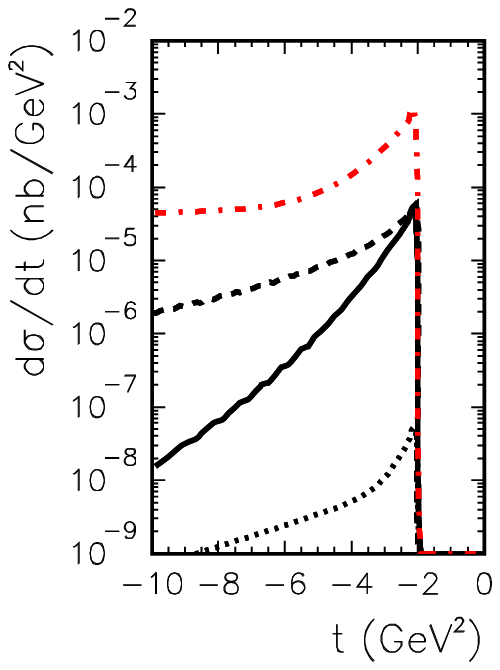}
\includegraphics[width=6cm]{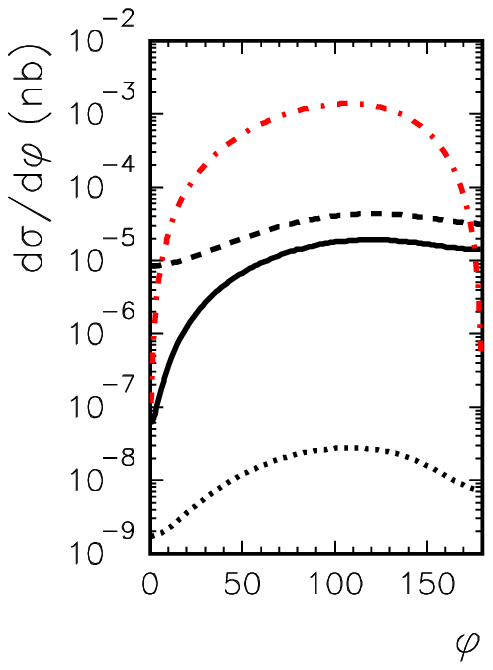} \\
\includegraphics[width=6cm]{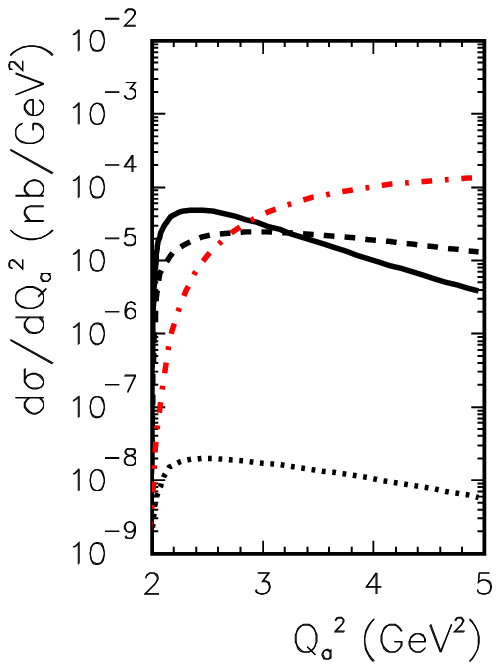}
\includegraphics[width=6cm]{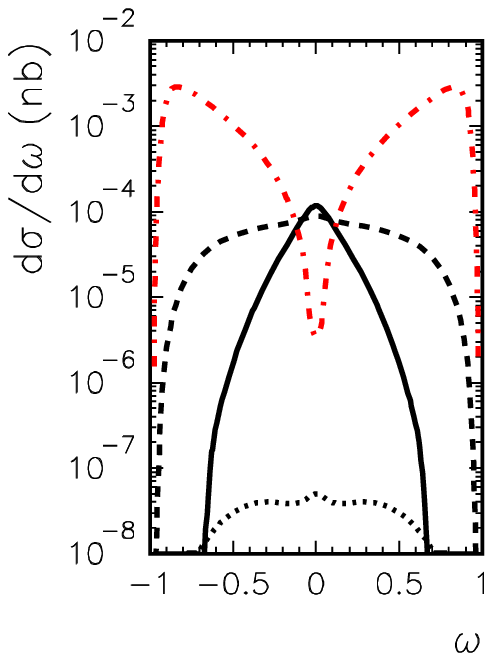}
\caption{Several distributions for production of $f_1(1285)$
in double-tagging mode with $Q_1^2, Q_2^2 > 2 GeV^2$.
The solid line is for LS, the dahed line for NQM, the dotted line for OPV
and the dash-dotted line for RS vertices.
}
\label{fig:distributions_for_DT}
\end{figure}

Now we wish to show several differential distributions for 
the double-tagged mode. In Fig.\ref{fig:distributions_for_DT}
we show distributions in rapidity and transverse momentum of
$f_1(1285)$, $t_1$ or $t_2$, azimuthal angle between outgoing electrons,
averaged virtuality
\begin{equation}
Q_a^2 = (Q_1^2 + Q_2^2)/2
\end{equation}
and the asymmetry parameter 
\begin{equation}
\omega = \frac{Q_1^2 - Q_2^2}{Q_1^2 + Q_2^2} \; .
\end{equation}
The Bose symmetry requires that:
\begin{equation}
\frac{d \sigma}{d \omega}(\omega) = \frac{d \sigma}{d \omega}(-\omega)
\; .
\end{equation}

Quite different distributions are obtained for the different vertices
used recently in the literature. Especially interesting are distribution
in relative azimuthal angle between outgoing electrons and distribution
in virtuality asymmetry $\omega$. For the RS2019 vertex \cite{RS2019} 
the vanishing of the cross section for $\omega$ = 0 is a consequence of 
the asymmetric form factor which goes to 0 for $Q_1^2 = Q_2^2$.
With the RS2019 vertex axial vector mesons do not contribute to
the hyperfine splitting of muonic atoms. 
It is obvious that the DT measurements of distributions
shown in Fig.\ref{fig:distributions_for_DT} would provide 
strong limitations on the vertices used in calculating
fundamental quantities such as muon anomalous magnetic moment $a_{\mu}$ 
and/or hyperfine splitting of muonic hydrogen.

%-------------------------------------------------------------
%--------------------------------------------------------------

%--------------------------
\section{Conclusions}
%--------------------------

In this paper the results of calculations of cross sections
and differential distributions for the $e^+ e^- \to e^+ e^- f_1(1285)$
have been performed using different $\gamma^* \gamma^* \to f_1(1285)$ 
couplings known from the literature. These couplings were used
previously to calculate hadronic light-by-light axial meson
contributions to anomalous magnetic moment of muon as well as for
hyperfine splitting of the muon hydrogen.

We have presented predictions relevant for future double-tagged 
experiments for Belle II. The results strongly depend on the details
of calculation (type of tensorial coupling and/or form factors used).
The form factor cannot be reliably calculated at present.
We have presented several diferential distributions in photon
virtualities, transverse momentum of $f_1(1285)$, 
distribution in azimuthal angle between outgoing electron and positron
and so-called asymmetry of virtualities ($\omega$). Especially the 
latter observable (asymmetry) seems promissing for verifying 
the quite different models of the $\gamma^* \gamma^* AV$ coupling.
The results strongly depend on details of the coupling(s).
The double tagged measurement would therefore be very valueable
to constrain the couplings and form factors and in a consequence
would help to decrease uncertainties of their contribution
to anomalous magnetic moment of muon and hyperfine splitting of muonic
hydrogen.

Both $\eta \pi^+ \pi^-$ (as in \cite{L3_f1}) as well as 
$\pi^+ \pi^- \pi^+ \pi^-$ (used recently at the LHC
\cite{Bols_master_thesis}) channels could be applied experimentaly 
to identify the $f_1(1285)$ meson.
The $\eta \pi^+ \pi^-$ option is dengerous as there is another
meson close by which decays to the same decay channel:
$\eta(1295) \to \eta \pi \pi$ \cite{Achasov_private}.
This meson may be also abundantly produced in $\gamma \gamma$ fusion
as $\Gamma_{\eta' \to \gamma \gamma}$ = 4.27 keV \cite{AAMN1999}.
$f_1(1285) \to \rho^0 \gamma$ with BR = 5.3 \% \cite{PDG} would
be another possible choice.
The decays of light axial vector mesons were discussed e.g. in 
\cite{RPO2004,RHO2007,CPLM2015}.

In the present paper we concentrated on production of $f_1(1285)$ meson.
A similar analysis could be performed for other axial-vector mesons such
as $a_1(1260)$ or $f_1(1420)$.
Then coupling constants and some form factors must be changed in 
the calculation. On the experimental side, decay channels specific
for a given meson must be selected.

The production of isoscalar axial-vector mesons is very interesting
also in the context of central exclusive processes $p p \to p p f_1$.
There the unknown ingredient is pomeron-pomeron-$f_1$ vertex.
This will be discussed elsewhere \cite{LNS2020}.

\vskip+5mm
{\bf Acknowledgments}\\
I am indebted to Wolfgang Sch\"afer for collaboration on quarkonium 
production in photon-photon collisions and Piotr Lebiedowicz
for collaboration on diffractive production of $f_1$ meson. 
The discussion with Anton Rebhan, Josef Leutgeb, Alexander Osipov, 
Pablo Roig and Pablo Sanchez-Puertas and Sasha Dorokhov on 
$\gamma^* \gamma^* \to f_1(1285)$ vertices and with Alexander Rudenko 
about $f_1(1285) \to e^+ e^-$ is acknowleged.
The symmetry relations were discussed with Otto Nachtmann.
The decays of $f_1$ and related difficulties were discussed
with Nikolay Achasov.
A possibility of a measurement at Belle II was discussed with
Sadaharu Uehara.
This study was partially supported by the Polish National Science Center
grant UMO-2018/31/B/ST2/03537 and by the Center for Innovation and
Transfer of Natural Sciences and Engineering Knowledge in Rzesz{\'o}w.

%----------------------------------------------------------------------------

%----------------------------------------------------------------------------

\end{document}